\documentclass[letter]{aa}
\usepackage{graphicx}
\usepackage{txfonts}
\newcommand{\kms}{\mbox{km~s$^{-1}$}}
\newcommand{\msun}{\mbox{$\mathcal{M}_{\odot}$}}

\begin{document}
  \title{Molecular Gas in NUclei of GAlaxies (NUGA):\\ VI.\ Detection
         of a molecular gas disk/torus via HCN in the Seyfert~2 galaxy
         NGC~6951?\thanks{Based on observations carried out with the
         IRAM Plateau de Bure Interferometer (PdBI). IRAM is supported
         by INSU/CNRS (France), MPG (Germany) and IGN (Spain).}}

\titlerunning{HCN(1--0) emission in NGC6951}
\authorrunning{Krips et al.}


   \author{M.\ Krips 
          \inst{1}
          \and
          R.\ Neri\inst{2}
          \and
	  S. Garc\'{\i}a-Burillo\inst{3}
          \and
	  F.\ Combes\inst{4}
          \and
	  E.\ Schinnerer\inst{5}
          \and
	  A.J.\ Baker\inst{6}
          \and
	  A.\ Eckart\inst{7}
          \and
	  F.\ Boone\inst{5}
	  \and
	  L.\ Hunt\inst{8}	  
          \and
	  S.\ Leon\inst{9}
	  \and
	  L.J.\ Tacconi\inst{10}
         }

   \offprints{M. Krips}

   \institute{Harvard-Smithsonian Center for Astrophysics, SMA project, 
              645 N A`ohoku Pl., Hilo, HI,96720, USA;
              \email{mkrips@cfa.harvard.edu} 
              \and 
              Institut de Radio Astronomie Millim\'etrique (IRAM),
              Saint Martin d'H\`eres, F-38406, France 
              \and 
              Observatorio Astron\'omico Nacional (OAN) -
              Observatorio de Madrid, C/ Alfonso XII 3, 28014 Madrid, Spain
              \and 
	      Observatoire de Paris, LERMA, 61 Av. de
	      l'Observatoire, 75014 Paris, France
              \and 
              Max-Planck-Institut f\"ur Astronomie, K\"onigstuhl
              17, 69117 Heidelberg, Germany 
              \and 
              Department of Physics and Astronomy, Rutgers, the State
	      University of NJ, 136 Frelinghuysen Rd., Piscataway, NJ
	      08854-8019, USA
              \and 
              Universit\"at zu K\"oln, I.Physikalisches Institut,
              Z\"ulpicher Str. 77, 50937 K\"oln, Germany
              \and
              INAF-Istituto di Radioastronomia/Sezione Firenze
	      Largo E. Fermi 5, 50125 Firenze, Italy
	      \and 
              IRAM, Avenida Divina Pastora, 7, N\'ucleo Central,
              18012 Granada, Spain
	      \and
              Max-Planck-Institut f\"ur extraterrestrische Physik,
              Postfach 1312, 85741 Garching, Germany
	      }
   \date{}

 
  \abstract
%
    {Several studies of nearby active galaxies indicate significantly
    higher HCN-to-CO intensity ratios in AGN (e.g., NGC~1068) than in
    starburst (e.g., M82) environments. HCN enhancement can be caused
    by many different effects, such as higher gas densities and/or
    temperatures, UV/X-ray radiation, and non-collisional
    excitation. As active galaxies often exhibit intense circumnuclear
    star formation, high angular resolution/high sensitivity
    observations are of paramount importance to disentangling the
    influence of star formation from that of nuclear activity on the
    chemistry of the surrounding molecular gas. The tight relation of
    HCN enhancement and nuclear activity may qualify HCN as an ideal
    tracer of molecular gas close to the AGN, providing complementary
    and additional information to that gained via CO.}
%
   {NGC~6951 houses nuclear and starburst activity, making it an ideal
   testbed in which to study the effects of different excitation
   conditions on the molecular gas. Previous lower angular
   resolution/sensitivity observations of HCN(1--0) carried out with
   the Nobeyama Millimeter array by Kohno et al.\ (1999a) led to the
   detection of the starburst ring but no central emission has been
   found. Our aim was to search for nuclear HCN emission and, if
   successful, for differences of the gas properties of the starburst
   ring and the nucleus.}
%
%
   {We used the new A, B, C and D configurations of the IRAM PdBI
   array to observe HCN(1--0) in NGC~6951 at high angular resolution
   (1$''$$\equiv$96~pc) and sensitivity.}
%
%
   {We detect very compact ($\leq$50pc) HCN emission in the nucleus of
   NGC~6951, supporting previous hints of nuclear gas structure.  Our
   observations also reveal HCN emission in the starburst ring and
   resolve it into several peaks, leading to a higher coincidence
   between the HCN and CO distributions than previously reported by
   Kohno et al.\ (1999a).}
%
%
   {We find a significantly higher HCN-to-CO intensity ratio
   ($\geq$0.4) in the nucleus than in the starburst ring
   (0.02-0.05). As for NGC~1068, this might result from a higher HCN
   abundance in the centre due to an X-ray dominated gas chemistry,
   but a higher gas density/temperature or additional non-collisional
   excitation of HCN cannot be entirely ruled out, based on these
   observations. The compact HCN emission is associated with rotating
   gas in a circumnuclear disk/torus.}


   \maketitle
%

\begin{figure*}[!t]
\centering
\rotatebox{-90}{\resizebox{7.9cm}{!}{\includegraphics{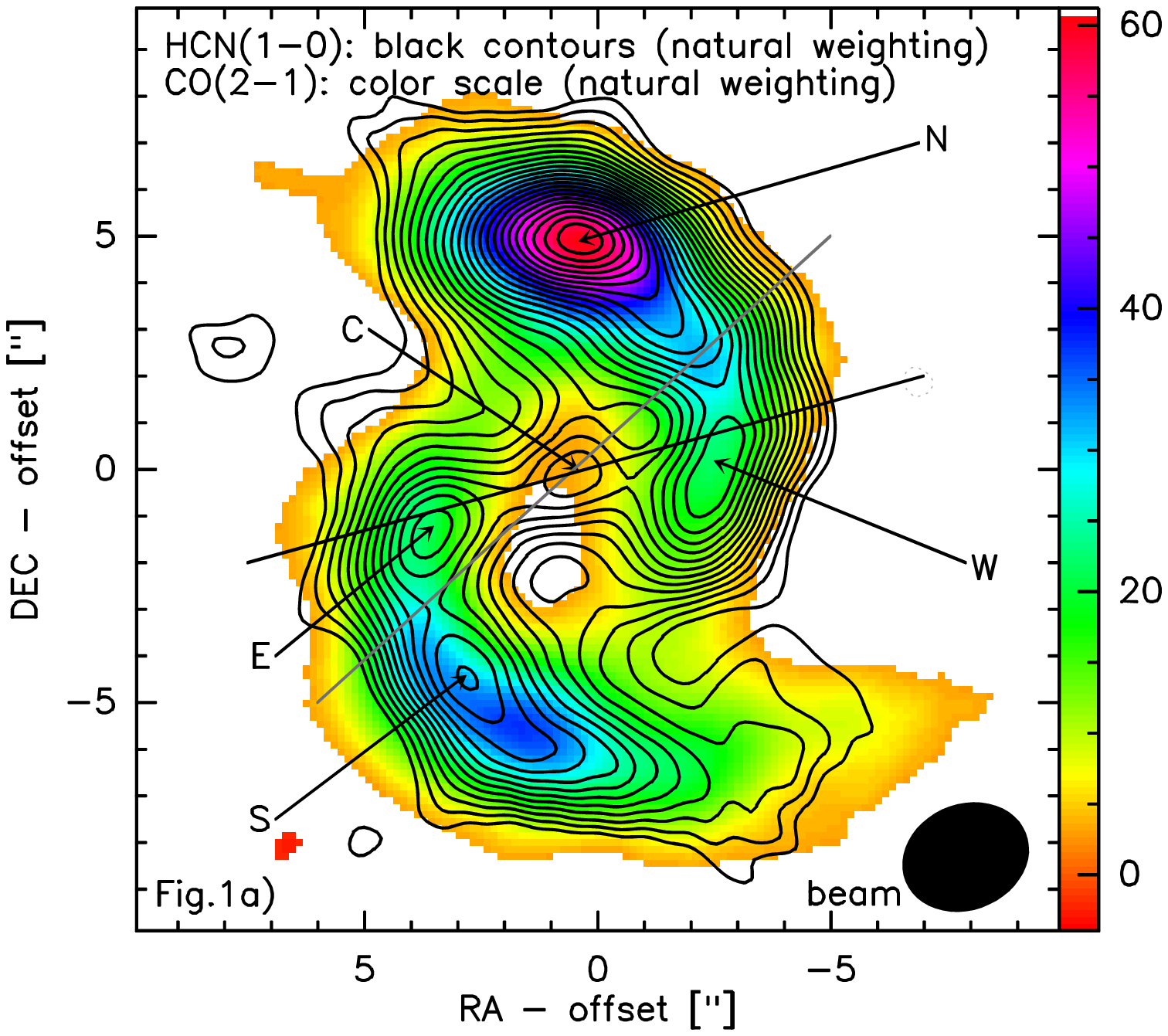}}}
\rotatebox{-90}{\resizebox{7.9cm}{!}{\includegraphics{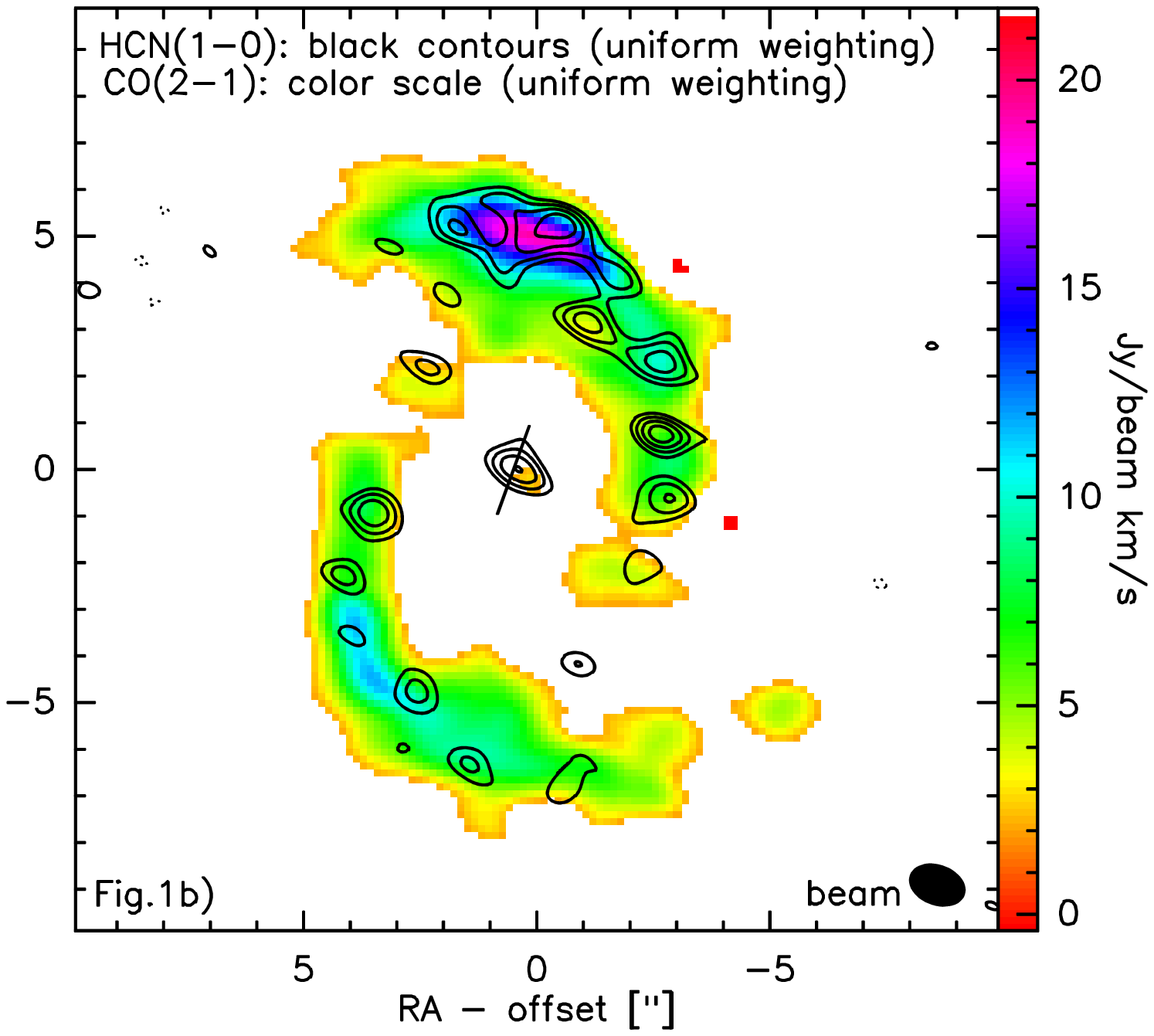}}}
\caption{Integrated HCN(1--0) emission ({\it black contours}) overlaid
on CO(2--1) ({\it color scale}; Schinnerer et al., in prep.,
Garc\'{\i}a-Burillo et al.\ 2005) in natural ({\it left}; 1a) and
uniform weighting ({\it right};1b); the CO and HCN emission have been
both integrated from $-$200\kms\ to $+$200\kms. We used a uv-taper for
CO to match the angular resolution of our HCN data which is by a
factor of $\sim$2 lower, and obtain identical beamsizes. Black
contours run from 3$\sigma$ to 23$\sigma$ ({\it right:} 7$\sigma$) in
steps of 1$\sigma$=0.06~Jy\,beam$^{-1}$\ \kms\, ({\it right:}
1$\sigma$=0.10~Jy\,beam$^{-1}$\ \kms); CO(2--1):
3$\sigma$=1.0~Jy\,beam$^{-1}$\ \kms ({\it right:}
3$\sigma$=1.3~Jy\,beam$^{-1}$\ \kms). The black line ({\it left})
indicates the major axis of the bar (PA=100$^\circ$) and the grey line
the major axis of the galaxy (PA=130$^\circ$). The black line ({\it
right}) represents the most extreme central velocity gradients
(PA=(160$\pm20$)$^\circ$). The (0,0) position is at $\alpha_{\rm
J2000}$=20$^{\rm h}$37$^{\rm m}$14.123$^{\rm s}$ and $\delta_{\rm
J2000}$=66$^\circ$06$'$20.09$''$ which is slightly different from the
phase centre of the observations.}
\label{co-hcn}
\end{figure*}

\section{Introduction}
Over the past decade, several studies (e.g., Tacconi et al.\ 1994,
Sternberg et al.\ 1994, Kohno et al.\ 1999a,b, 2000, 2001, 2005) have
shown that the HCN-to-CO intensity ratio ($\equiv$R$_{\rm HCN/CO}$)
can be significantly higher in active galactic nuclei (AGN; e.g.,
NGC~1068) than in starburst (SB; e.g., M82) or quiescent
regions. While surprisingly large R$_{\rm HCN/CO}$ (up to 1 or more)
have been observed in AGN (e.g., NGC\,1068, NGC\,1097, NGC\,5194;
Kohno et al.\ 1999a, Tacconi et al.\ 1994, Usero et al.\ 2004), much
smaller R$_{\rm HCN/CO}$ ($<$0.3) are detected in pure SB or composite
(AGN+SB) galaxies (e.g., Arp220, M82, NGC~6951; Gao \& Solomon
2004a,b, Nguyen-Q-Rieu et al.\ 1992). Inactive galaxies have even
lower ratios of R$_{\rm HCN/CO}$$<$0.1.  Many different effects can
contribute to increased R$_{\rm HCN/CO}$ in active environments
including higher gas opacities/densities and/or temperatures,
non-standard molecular abundances caused by strong UV/X-ray radiation
fields or additional non-collisional excitation such as IR pumping
through UV/X-ray heated dust. Gao \& Solomon (2004a) have ruled out
the latter scenario for large scale HCN emission and the lack of any
clear correlation between the hard X-ray and MIR luminosity in AGN
(Lutz et al.\ 2004) reduces the significances of IR pumping also at
small scales. However, Usero et al.\ (2004) present strong evidence in
the case of NGC~1068 that the nuclear gas chemistry is dominated by
X-ray radiation from the AGN yielding significantly different
molecular abundances than in SB or quiescent environments (Lepp \&
Dalgarno 1996; Maloney et al.\ 1996). Recent IRAM 30m observations of
several HCN transitions in a sample of 12 nearby active galaxies also
seem to support a significantly higher HCN abundance in AGN than in SB
or inactive environments, rather than a pure density/temperature or
non-collisional excitation effect (Krips et al., in prep.). If true,
this has a severe impact on the interpretation of R$_{\rm HCN/CO}$ as
a measure of the dense to total molecular gas mass fraction in active
galaxies (e.g., Gao \& Solomon et al.\ 2004a,b) as discussed in
Graci\'a-Carpio et al.\ (2006). The study of nearby active galaxies
also reveals the limitations of R$_{\rm HCN/CO}$ as a unique
diagnostic in distant sources whose starburst and AGN components
cannot be separated.

\begin{table}[!]
\label{tab1}      
\centering                          
\begin{tabular}{c c c c c}       
\hline\hline                 
Component & peak flux & FWHM  &  I$_{\rm HCN(1-0)}$ &  I$_{\rm CO(2-1)}$$^a$ \\   
         & (mJy\,beam$^{-1}$) & (\kms) & \multicolumn{2}{c}{(Jy\,beam$^{-1}$~\kms)}  \\
\hline                        
  N & 11$\pm$2.0  & 150$\pm$6  &  1.8$\pm$0.2 & 60$\pm$6 \\      
  W & 10$\pm$1.0  & 100$\pm$8  &  1.1$\pm$0.1 & 22$\pm$2 \\      
  E & 12$\pm$1.0  &  70$\pm$5  &  1.0$\pm$0.1 & 23$\pm$2 \\      
  S & 13$\pm$1.0  &  60$\pm$4  &  0.9$\pm$0.1 & 38$\pm$4 \\      
  C$^b$   & 4.8$\pm$0.5 & 170$\pm$20 &  0.9$\pm$0.1 & 2.4$\pm$0.2 \\
\hline
  Total flux$^c$         & 36.0$\pm$4.0$^d$  & 320$\pm$14 & 12.0$\pm$1.0$^e$ & 470$\pm$50$^e$ \\
\hline                                   
\end{tabular}
\caption{HCN(1--0) line parameters, obtained at the various peak
(i.e., not spatially integrated) by fitting a Gaussian profile to the
(naturally weighted) data. Errors include uncertainties of the fit and
calibration. $^a$ from Schinnerer et al., in prep. $^b$ from uniformly
weighted maps to avoid contamination by the ring emission. $^c$
spatially integrated over the entire area of $\pm$9$''$; $^d$ in mJy;
$^e$ in Jy~\kms.}
\end{table}

NGC~6951 is an active galaxy of Hubble type SAB(rs)bc at a distance of
24~Mpc (Tully 1988); its active nucleus is classified as a transition
object between a LINER and a type 2 Seyfert (P\'erez et al.\ 2000). In
addition to its AGN, NGC6951 also exhibits a pronounced SB ring at a
radius of 5$''$ ($\equiv$480~pc) in H$\alpha$ (Marquez \& Moles 1993,
Wozniak et al.\ 1995, Rozas et al.\ 1996; Gonzalez-Delgado \& Perez
1997, Perez et al.\ 2000) and radio emission (Vila et al.\ 1990,
Saikia et al.\ 1994 \& 2002). Strong CO and HCN emission is associated
with the SB ring (e.g., Kohno et al.\ 1999a, Garc\'{\i}a-Burillo et
al.\ 2005) while almost no emission had been hitherto found in the
centre of NGC~6951. Only recently have high angular resolution/high
sensitivity PdBI observations revealed faint CO(2--1) emission in the
central 0.5$''$ (Garc\'{\i}a-Burillo et al.\ 2005; Schinnerer et al.,
in prep.). The latter observations are part of the PdBI NU(clei
of)GA(laxies) project (e.g., Garc\'{\i}a-Burillo et al.\ 2003). We
observed NGC~6951 in HCN(1--0) to search for nuclear emission and
assess differences between the SB ring and the AGN; the results of
these observations are presented here.

\section{Observations}
\label{obs}
NGC~6951 has been observed at 3mm and 1mm using all six antennae of
the IRAM PdBI in the new A, B, C and D configurations (see
http://www.iram.fr for telescope parameters) during January, February
(A+B), April and May (C+D) 2006 . The phase reference centre was set
to $\alpha_{\rm J2000}=$20$^{\rm h}$37$^{\rm m}$14.470$^{\rm s}$ and
$\delta_{\rm J2000}=$66$^\circ$06$'$19.70$''$. The 3mm receivers were
tuned to the frequency of HCN(1--0) shifted to the LSR velocity of
$v_{\rm LSR}=$1424\kms, while the 1mm receivers were set to the
$^{13}$CO(2--1) line. The 1mm data will be discussed in a separate
paper (Krips et al., in prep.). Upper and lower sidebands at each
frequency had 580 MHz bandwidth and 1.25 MHz resolution.  Weather
conditions were good throughout the observations with a water column
varying between 4mm and 10mm and SSB system temperatures of 100-150~K
(A+B) and 100-200~K (C+D) at 3mm.  3C273, 3C454.3, 2200+420 and
1749+096 were used as bandpass calibrators while 1928+738 and 2037+511
were used as gain calibrators. Fluxes have been calibrated with MWC349
and checked on 3C273, 1928+738 and 2037+511 with the flux monitoring
program of the IRAM PdBI, resulting in an accuracy of $\sim$10\% at
3mm.  The data were calibrated, mapped and analyzed using the standard
IRAM GILDAS programs CLIC and MAPPING. Using uniform weighting, the
synthesized beam size is 1.25$''\times$0.87$''$ at a position angle
(PA) of 72$^\circ$ at 3mm; natural weighting results in 2.78$''\times$
2.27$''$ at a PA of 112$^\circ$. We reach an rms noise of
$\sim$0.8~mJy\,beam$^{-1}$ ($\sim$1.2~mJy\,beam$^{-1}$) at a spectral
resolution of 5~MHz ($\equiv$17\kms) at 3.4~mm for natural (uniform)
weighting.


\begin{figure}[!t]
\centering
\resizebox{8.3cm}{!}{\rotatebox{-90}{\includegraphics{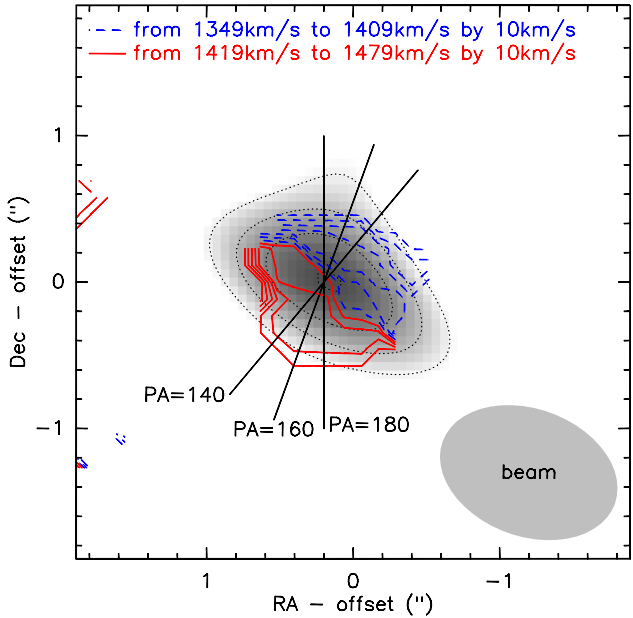}}}
\caption{{\rm Iso-velocity map of HCN(1--0) ({\it solid red \& dashed
blue lines}) overlaid to the integrated HCN(1--0) emission ({\it grey
scale \& grey contours} (same as in Fig.~\ref{co-hcn})). Contours run
from v$_{\rm LSR}$=1349~\kms\, ({\it dashed blue}) to 1479~\kms\,
({\it solid red}) in steps of 10~\kms around the dynamical center
($\sim$10~\kms\, off from the systemic velocity). The black lines
indicate the PA (=160$\pm$20$^\circ$) of the velocity gradient.}}
\label{hcn-slice}
\end{figure}

\section{Results \& Discussion}

\subsection{Starburst ring}
HCN emission is clearly detected along the starburst ring, consistent
with Kohno et al.\ (1999a). The total flux observed with the PdBI is
lower by a factor of $\sim$2-3 than the flux obtained with the IRAM
30m telescope (Krips et al., in prep.), i.e., some of the HCN emission
is resolved out by the PdBI. The two main maxima, which are seen in
HCN by Kohno et al.\ (1999a) at an angular resolution of $\sim$4.5$''$
and peak at different positions than CO, split up into several
sub-maxima (N,W,E,S; Fig.~\ref{co-hcn}) at the higher angular
resolution ($<$3$''$) of the PdBI. The HCN peaks N and S are
consistent in position with those in CO (Fig.~\ref{co-hcn};
Garc\'{\i}a-Burillo et al.\ 2005). Additional HCN peaks are found to
the west (W) and east (E) (see Fig.~\ref{co-hcn}). The W-to-N and
E-to-S peak ratios are higher in HCN ($\sim$0.6-1.0) than in CO
($\sim$0.4-0.6; Table~\ref{tab1}) explaining why the positions of the
merged peaks in Kohno et al.\ (1999a) do not agree between HCN and
CO. The HCN-to-CO ratios ($\equiv$R$_{\rm HCN/CO}$=I$_{\rm
HCN(1-0)}$/I$_{\rm CO(1-0)}$; with I$\equiv$velocity integrated
intensity) thus also differ between the S/N and W/E peaks. While
R$_{\rm HCN/CO}$ is found to be roughly $\sim$0.03 at S and N
(assuming I$_{\rm CO(2-1)}$/I$_{\rm CO(1-0)}$=1; Table~\ref{tab1}), we
estimate R$_{\rm HCN/CO}\simeq0.05$ at E and W. This might indicate a
variation of the gas density/temperature along the ring. The HCN
kinematics in the ring agree well with those seen in CO and are hence
not further discussed in this letter.

\subsection{Central emission}
HCN(1--0) emission is detected in the central $\sim$1$''$
($\simeq$100~pc; $\equiv$C) because of the higher sensitivity and
angular resolution of our data compared to the one obtained by Kohno
et al.\ (1999a), confirming previous hints found in CO(2--1)
(Garc\'{\i}a-Burillo et al.\ 2005; Schinnerer et al., in prep.). The
nuclear HCN component C appears to be compact and unresolved in the
PdBI beam. Assuming the ``HCN conversion'' factor of X$_{\rm
HCN}$(=M$_{\rm HCN}$(H$_2$)/L$_{\rm HCN}$)=20$_{-10}^{+30}$~\msun\ (K\
\kms\ pc$^2$)$^{-1}$ from Solomon et al.\ (1992), we find a central
dense gas mass of M$_{\rm
gas}$(=M(H$_2$+He))$\approx$(2-10)$\cdot$10$^7$\msun\ which is a
factor of $\sim$3-17 higher than the mass of
$\sim$6$\cdot$10$^6$\msun\ derived from the CO(2--1) line (assuming
I$_{\rm CO(2-1)}$/I$_{\rm CO(1-0)}$=1 (Table~\ref{tab1}; see also
Garc\'{\i}a-Burillo et al.\ 2005). This large discrepancy cannot be
solely due to the large uncertainties of the conversion factors but
might further indicate non-standard gas conditions in the AGN. The
HCN-to-CO ratio of the velocity integrated line flux amounts to
$\geq$0.4 (Table~\ref{tab1}; I$_{\rm CO(2-1)}$/I$_{\rm
CO(1-0)}$=1). This is a factor of $\sim$4-8 larger than in the
starburst ring and the value of 0.09 for the nucleus published by
Kohno et al.\ (1999a) indicating that the nuclear HCN is significantly
enhanced in NGC~6951.  Thus, unlike Kohno et al.\ (1999a), we conclude
that the nuclear gas properties in NGC~6951 are {\it not} that
different from those in the Seyfert galaxies NGC~1068 or M51, both of
which have similarly high central HCN-to-CO ratios. The molecular gas
chemistry in the AGN influenced regions of NGC~1068 and NGC~5194 is
dominated by X-ray radiation (NGC~1068: e.g., Usero et al.\ 2004; M51:
e.g., Matsushita et al.\ 1998) suggesting a similar scenario in
NGC~6951. Thus, the central dense gas mass derived via HCN has to be
considered an upper limit.

The central HCN emission shows the steepest velocity gradient at
PA$\approx$160$^\circ$ with a velocity range of $\pm$70~\kms\ around
the dynamical center over a radius of $\sim$0.5$''$
(Fig.~\ref{hcn-slice}), eventually indicating a gas rotation in a
circumnuclear disk or torus. Even if accounting for the uncertainties
of the PA, estimated to be $\pm$20$^\circ$, the central kinematic axis
is significantly different from the major axis of the galaxy
(PA=135$^\circ$) and of the bar (PA=100$^\circ$). This non-alignment
might be caused by a different inclination of the central gas
disk/torus than of the galaxy, or, alternatively, by non-circular
velocities or a warp. However, the central velocity gradient allows us
to give a crude estimate of the enclosed dynamical mass. Assuming a
radius of $\sim$50~pc and a velocity range of $\pm$70\kms, we find
M$_{\rm dyn}$=5$\cdot$10$^7\cdot(\sin{i})^{-2}$\msun, with
$i$$\equiv$inclination of the disk. This is similar to the gas mass of
0.6-8$\cdot$10$^7$\msun, suggesting that the disk is seen closer to
face-on than to edge-on. Assuming further the inclination of the
galaxy ($\approx$40$^\circ$; Garc\'{\i}a-Burillo et al.\ 2005) as
rough estimate for the one of the central gas disk, the enclosed
dynamical mass increases to $\sim$2$\cdot$10$^8$\msun. This is of the
order of the nuclear black hole mass of $\sim$2$\cdot$10$^8$\msun,
estimated via the stellar velocity dispersion of the bulge
($\sigma_s$=232\kms from Ho \& Ulvestad 2001; see also Gebhardt et
al.\ 2000). However, within a radius of 50~pc, stars might still
contribute a significant fraction to the dynamical mass as well
pointing towards an even lower inclination of the nuclear disk than
assumed.

\section{Conclusions}
We have presented high angular resolution/sensitivity observations of
the HCN(1--0) emission in NGC~6951 in this letter which exploit the
new ABCD configurations at the IRAM PdBI. The main results can be
summarized as follows:

   \begin{enumerate}
      \item The two main HCN peaks in the ring seen by Kohno et al.\
            (1999a) split up into several sub-maxima which coincide
            with those found in CO(2--1) by Garc\'{\i}a-Burillo et
            al.\ (2005) and Schinnerer et al.\ (in prep.). However,
            the ratios of peaks within the ring differ between HCN and
            CO, probably indicating different gas
            densities/temperatures along the ring.
      \item We clearly detect compact HCN(1--0) emission in the
            nucleus of NGC~6951 tracing a dense molecular gas mass in
            the range of $\sim$0.6-6$\times$10$^7$\msun. A position
            velocity cut taken along PA=(160$\pm$20)$^\circ$,
            coinciding neither with the galaxy nor the bar major axis,
            yields the steepest velocity gradient in the nucleus. The
            gradient and the compactness of this central HCN emission
            suggest that it might arise in a rotating circumnuclear
            gas disk or torus with a radius of $\lesssim$50~pc.
      \item In contrast to Kohno et al.\ (1999a), we find
            significantly different HCN-to-CO line ratios in the
            starburst ring and the nuclear emission. While the
            starburst ring shows typical R$_{\rm HCN/CO}$=0.02-0.05,
            HCN is significantly enhanced in the centre with R$_{\rm
            HCN/CO}\geq$0.4. Either the molecular gas in the nucleus
            is denser and/or hotter than in the starburst ring,
            increasing R$_{\rm HCN/CO}$, or the gas chemistry in the
            nucleus of NGC~6951 is dominated by X-ray radiation from
            the AGN, producing a higher nuclear HCN abundance and thus
            a higher R$_{\rm HCN/CO}$ similar to the center of
            NGC~1068.
   \end{enumerate}


\end{document}